\documentclass{iau_FM}
\usepackage{graphicx}

\title[IAU-FM6.-- Angular momentum -- Conference summary] 
{Angular momentum -- Conference summary}

\author[F. Combes]   
{Francoise Combes$^1$}

\affiliation{$^1$Observatoire de Paris, LERMA, College de France, CNRS, PSL Univ., Sorbonne Univ.,
F-75014, Paris, France \\ email: {\tt francoise.combes@obspm.fr}}

\pubyear{2018}
\jname{Astronomy in Focus, Volume 1}
\editors{Danail Obreschkow, ed.}
\begin{document}

\maketitle

\begin{abstract}
Angular momentum (AM) is a key parameter to understand galaxy formation and
evolution. AM originates in tidal torques between proto-structures
at turn around, and from this the specific AM is expected to scale as 
a power-law of slope 2/3 with mass. However, subsequent evolution
re-shuffles this through matter accretion from filaments, mergers, 
star formation and feedback, secular evolution and AM exchange between
baryons and dark matter. Outer parts of galaxies are essential to study
since they retain most of the AM and the diagnostics of the evolution.
Galaxy IFU surveys have recently provided a wealth of kinematical 
information in the local universe. In the future, we can expect more statistics
in the outer parts, and evolution at high z, including atomic gas with SKA.
 
\end{abstract}


This focus meeting has emphazised the high importance of angular momentum,
to understand the formation and evolution of galaxies. It is now used
extensively, given the progress of IFU instruments and large galaxy surveys.

Given these recent developments, it is difficult to imagine the debate that
was occurring only 60 years ago, about the origin of the angular momentum
of galaxies. The theory was first proposed by C. von Weizs\"acker that
galaxies were originating in large eddies of cosmic turbulence. This theory
was followed by many people like G. Gamow, V. Rubin, his student or J. Oort.

Jim Peebles convinced Jan Oort that turbulence was irrelevant, that gravity
and tidal torques could create the right amount of angular momentum (AM). For that
he computed the torques with N-body simulations (N=90) and showed that
the un-dimensional value of the AM $\lambda= \frac{J |E|^{1/2}}{G M^{5/2}} \sim 0.1$,
in agreement from analytical estimations.

Since then, dark matter  has been introduced, the problem is more complex,
since we observe only the angular momentum of the baryons, which has to be related
to the dark matter one.
How are these acquired, how do they exchange?

The first cosmological simulations with baryons and dark matter,
pointed out a serious problem, called  the AM catastrophy: the baryons
were losing their angular momentum through dynamical friction in mergers
in favor of the dark matter, and were accumulating in very small disks at
the bottom of the potential wells. Thanks to the feedback, and also the
increase in spatial resolution of the simulations (lowering the effects of friction),
the AM catastrophy is now limited (e.g. Obreja, Pedrosa and others, this meeting).

\section{The ``Fall'' relation}

In their pioneering study, Fall \& Efstathiou (1980) take into account baryons and
dark matter, which was only made of old stars at this epoch.  Fall (1983) considers
several scenarios of AM, mass or energy conservation, and concludes that the best
scenario fitting the observations is that of baryonic mass M and AM conserved, while energy is dissipated.
In this case, the specific angular momentum, i.e.  j= J/M is a power-law function of mass,
with slope 2/3. Several parallel lines can be traced, with the same slope in the logj-logM diagram,
the highest one is for very late disk galaxies (Sc), while the early-type galaxies (ETG) 
fall below, due to their high velocity dispersion and low rotation (low V/$\sigma$).
When only dark matter halos are concerned, the Virial relation combined with
the hypothesis that all halos at any mass are formed out of a constant volumic density,
leads to the power-law relation with slope 2/3.

Thirty years later Romanowsky \& Fall (2012), and Fall \& Romanowsky (2013) follow up
using the much better determined AM and the much larger statistics provided by modern
galaxy surveys. They show that the specific j can be used as a new classification scheme
for galaxies, since all the Hubble sequence can be retrieved through parallel lines of
2/3 slopes in the logj-logM baryonic diagram.
Many other versions of this diagram and classification were published
(Obreschkow \& Glazebrook, 2014; Cortese et al., 2016; Posti et al., 2018;  Sweet et al., 2018).

All these studies led to consider a third parameter in the AM scaling relation:
the relation can be viewed in a 3-dimension space, where the third axis is the bulge to total mass
ratio B/T (Fall \& Romanowsky 2018, also Obreschkow  \& Glazebrook 2014). The scaling relation
M-j-B/T can then be retrieved from the well known Tully-Fisher relation for spirals, and fundamental plane
for early-type galaxies. together with a structure relation (for instance the Freeman's relation
M $\propto$ R$^2$ for high-surface brightness spirals).

\section{$\Lambda$CDM hydro numerical simulations}

In the recent years, there has been a burst of simulation papers, interested
in following angular momentum, as described by Susana Pedrosa in her review
(Pedrosa \& Tissera 2015, Genel et al. 2015, Teklu et al. 2015, Obreja et al. 2016, 2018,
Lagos et al.2018).
Although the most realistic simulations, including star formation and feedback,
have solved the AM catastrophy (through the effect of feedback and higher spatial resolution),
they have revealed that the scaling relations of specific AM (j) versus baryonic mass
are flatter than those observed. The various galaxies follow parallel lines
in the  logj-logM baryonic diagram, with the B/T parameter increasing towards the bottom right,
but the slope of the lines are nearly 1/3.

Although the stellar feedback helps to solve the AM catastrophy, it also excessively
thickens galaxy disks. Simulations still predict too massive bulges, and feedback is not
sufficient to produce the large number of observed bulgeless galaxies. 

James Bullock remarked that very different results (especially in density and
temperature) can be obtained in general in cosmological simulations
when using different codes, different algorithms (Eulerian, Lagrangian),
different resolutions, different recipes for star formation and feedback.
However, the results on angular momentum, either of stars (j$_*$) or gas (j$_{gas}$)
are  converging! 

Due to dissipation, gaseous filaments are much thinner than dark matter filaments.
This means that even before matter enters into galaxies, the specific AM of baryons is
3 times higher than the specific AM of dark matter. This changes the initial conditions
in general adopted in semi-analytical models, where baryons and dark matter
are assumed to have gained the same specific j through tidal torques.
The virial radius R$_V$ changes a lot with time, it increases by a factor $\sim$ 3 from
z=1 to z=0. Since j $\propto$ $\lambda$ R$_V$, it is still possible that
the size of baryonic disks are the same at the end. The final j will depend on
the AM of the gas accreted in the mean time.

The size ratio between the stellar and dark matter components decreases with time for low M,
this was not reproduced before by the semi-analytical models. Now abundance matching is
considering sizes, as Rachel Somerville showed in her talk.

\begin{figure}[h]
\begin{center}
  \includegraphics[width=0.9\textwidth]{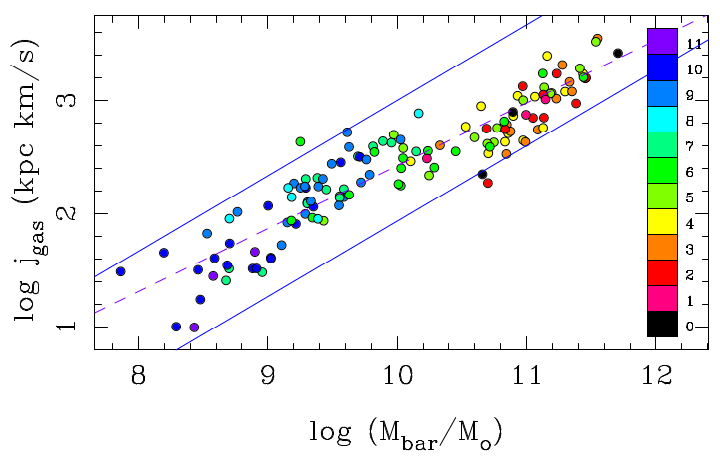}
  \caption{ The specific gas angular momentum j$_{gas}\propto$ R$_d$ V$_[flat$, versus the 
    baryonic mass M$_{bar}$= M$_*$ +M(HI), 
    from the 175 spiral galaxies of the SPARC sample of
    Lelli et al. (2016). The atomic gas
    is rotating maximally (negligible velocity dispersion), and the diagram should follow
    the upper envelope with a slope 2/3. In fact, the best fit has a slope of 0.55. The colour
    indicates the galaxy type, 0 being a lenticular, then Sa, Sab .. 9 is Sm, 10 Im and 11 BCD.}
   \label{fig1}
\end{center}
\end{figure}

\section{Why such a scaling relation?}

The observation of the logj$_*$ - logM$_*$ scaling relations in parallel lines
with a slope 2/3 is not straightforward to interpret. The first predictions were
done with the total matter, and can be applied essentialy to the dark matter,
but it is not obvious why the stars would follow the same relation.

Posti et al. (2018a,b) have proposed some biased collapse scenario,
to explain why the baryons do not retain all their initial angular momentum.
However, the scenario must be rather contrived. Indeed, to derive from the dark matter relation
j$_{DM}$ = J$_{DM}$/ M$_{DM} \propto$ M$_{DM}^{2/3}$, the equivalent relation for stars,
j$_* \propto$ f$_j$ f$_*^{-2/3}$ M$_*^{2/3}$, we must assume that the product
f$_j$ f$_*^{-2/3}$ = cst, with f$_j$ = j$_*$/j$_{DM}$ and f$_*$ = M$_*$/M$_{DM}$.
This last ratio is the well known fraction of stellar mass in a galaxy,
which is much below the universal baryon fraction f$_b$=17\%. From abundance
matching, this function peaks for halos of the Milky Way mass, and then falls
steeply on each side by 2 orders of magnitude (e.g. Behroozi et al. 2010).
To interpret the AM observations, we should explain why the f$_j$ ratio has the same 
behaviour, more exactly f$_j \propto$ f$_*^{2/3}$. the biased collapse scenario
proposed by Posti et al. (2018b) requires that the outer parts of halos, rich in AM,
fail to accrete on the galaxy to form stars.
This requirement looks like a conspiracy!

May be the specific AM of baryons does not always follow the scaling relation
with slope 2/3. When dwarfs dominated by dark matter and 
gas are considered, the slope is more near 0.5, as shown in Figure \ref{fig1}.

\section{Exchanges of AM  -- Secular evolution}

During galaxy evolution, angular momentum is not frozen either in
the baryons or dark matter, but their fraction may vary.
AM can be exchanged through spiral arms within the disk, which produces
radial migration. Some density breaks in the radial distribution
of stars can be attributed to these processes (Athanassoula  2014,  Peschken et al. 2017).
Bars exchange AM with the dark  halo, enhancing the formation of bars,
which are waves of negative angular momentum. Bars can also be destroyed
through torquing the gas, which is driven to the center.

It is interesting to follow AM along cosmic filaments. Galaxies have special
orientations with respect to filaments: spirals have their spin parallel to them,
while ellipticals, coming from mergers of spirals, have their spin
perpendicular to them. 
The fraction of fast rotators (at least faster than the average) is increasing
with the distance to the filaments. Galaxy surveys begin to be able
to check all these predictions.
(Welker et al. 2014, Xiaohu Yang et al. 2018)

\section{Large complexity in AM evolution}

Shy Genel described a long long equation, supposed to control the evolution
of the angular momentum, and follow its evolution along
a galaxy life, with matter accretion and major mergers.
All parameters have to be taken into account, such as the stars formed
in situ, or ex-situ, the gas forming stars, and what happens during the feedback,
the new star formation from the gas lost, the gas accretion,
the minor mergers, the radial migration, the AM exchange with DM.
All this is far from the AM prediction from torques at turn-around,
and the scaling relation of  j $\propto$ M$^{2/3}$.

How can we explain this miracle?

First the envelope at high j applies to pure disks, with 100\% efficiency to
retain AM. This is relatively
obvious if material is almost in circular orbits:
this plays the role of an attractor (see the talk from Francesca Rizzo,
and Rizzo et al. 2018).
Then you depart progressively from this attractor, as soon as you
form bulges, spheroids, heating the stellar component, without the possibility of gas
 cooling.

\section{Apparent contradictions}

AM is a proxy for morphological types, as Fall \& Romanowsky (2013)
proposed. It is also well known that morphological types are
segregated by the densiy of environment (Dressler et al., 1980).
Spirals are dominating in the field, while their abundance decreases
at high galaxy density in favor of lenticulars and ellipticals. 
Michele Capellari (2016) in his review article proposes to apply
this segregation with density to fast and slow rotators, to replace
the spiral/elliptical classification.  And indeed, slow 
rotators are found at density  peaks in clusters and groups.

But in her talk, Jenny Greene claimed that there is no evidence of
environment effect on the AM of early-type galaxies (Greene et al., 2018).
This is obtained from many surveys (MASSIVE, SAMI, MANGA), and the
AM depends only on mass.

 Another issue when considering AM, is to know whether studies are extending enough in radius.
As described beautifully by Matthew Colless, we are witnessing 
a golden age for kinematical studies of galaxies, with integral field units (IFU) 
large surveys (Atlas3D, SAMI, CALIFA, MANGA etc..).
However, large numbers (thousands) of galaxies are observed only to Re, 
and hundreds to 2Re. In general you need HI surveys to reach the flat portion
of rotation curves, richer in AM.

In the optical, the kinematics of Globular Clusters (GC)
show that the spin and ellipticity increase in S0, while they
drop in Ellipticals with radius, as described with the 
SLUGGS survey by Jean Brodie (Brodie \& Romanowsky 2016). 
With Planetary Nebulae (PNe) Pulsoni et al. (2018) go much further in radius,
to 15-20 Re, where all the AM and signatures of the galaxy formation
subsist. There is a large diversity of situations for ETG.
Some slow rotators begin to rotate in the outer parts, and
among fast rotators, 70\% slowly rotate in the outer parts.

The transition radius between in-situ and ex-situ material
is $\propto$ 1/M$_*$: i.e. there is
more ex-situ material in massive galaxies, formed through mergers.
This is perfectly compatible with Illustris simulations
(Rodriguez-Gomez et al. 2016).

Lagos et al. (2018) have measured in 
detail through simulations how galaxies
gain and lose AM by matter accretion and mergers.
Dry mergers reduce specific j by 30\%, while
wet mergers inscrease j by 10\%.

\section{Atomic gas and dwarfs} 

As shown in Obreschkow et al. (2016) and in Murugeshan's talk,
the angular momentum has a large influence in the
stability of spiral galaxies and their HI gas fraction.
The stability criterion can be written as
q = j $\sigma_v$/GM $\propto$M$^{-1/3}$,
and the HI gas fraction f$_{atm}$ is AM-regulated and also $\propto$M$^{-1/3}$.
A related study by Romeo \& Mogotsi (2018) on stability and AM regulation 
includes the thickness of the stellar disk T$_*$, i.e. 
Q$_*\sim \sigma_v$ T$_*$.

In Chengalur's talk, another discrepancy between simulations and 
observations was revealed for dwarf galaxies: the specific AM of baryons j$_b$
increases below a baryonic mass of 10$^{9.1}$M$_\odot$, with respect
to the M$^{2/3}$ expected scaling relation (Kurapati et al. 2018).
For these dwarfs, disks become thicker due to star formation feedback,
and to the shallow potential well.
There is no dependency on large-scale
environment, so this is not due to possible accretion.
 Another explanation is that such dwarfs are dominated
by dark matter, therefore their observed rotational
velocity is much higher with respect to their visible mass (M$_{bar}$)
than for spiral of larger masses.

In FIRE simulations, dwarfs have very low rotational support:
the large SF feedback gives them a rounder shape
(El-Badry et al. 2018), and their specific j falls below
 the M$^{2/3}$ scaling relation.

\section{Perspectives}

May be all diagnostics of galaxy evolution are retained
 in the outer parts: accretion, ex-situ star formation, etc.
In that case PNe are the best tracers of AM and
evolution. It is of prime importance to  acquire more statistics,
for instance in the Hector IFS survey, 10$^5$ galaxies will
be obtained. Also other parameters must be followed, 
metallicity, stellar populations (see Kassin's talk).

With ELT and JWST, it will be possible to track the evolution with redshift.
We know already that galaxies become clumpy at $z>2$ and have much lower j$_*$.
While it is predicted that
j$_*\sim$ (1+z)$^{-1/2}$ (Obreschkow et al. 2015), F. Fraternali in his
talk found no evolution with z.

It is also paramount to study external accretion of gas, 
which contains a lot of AM, is at the origin of warps, etc.
HI maps are badly needed at intermediate and high z; in the future
SKA will provide a large number of these gas maps.

\end{document}